\documentclass[aps,superscriptaddress, bibliography, onecolumn, reprint, showpacs, footnoteinbib, groupedaddress]{revtex4-1}
\usepackage{graphicx}
\usepackage{amssymb}   
\usepackage{amsfonts}
\usepackage{amsmath}
\usepackage{dsfont}
\usepackage{natbib}
\usepackage{dcolumn}   
\usepackage{bm}        
\usepackage[mathscr]{eucal}
\usepackage[dvipsnames]{xcolor}
\usepackage{hyperref}
\usepackage[all]{hypcap} 
\usepackage{mathtools}
\usepackage{dsfont}
\usepackage{physics}
\usepackage{microtype}
\usepackage{setspace}

\usepackage{newtxtext}
\usepackage{newtxmath}

\DeclareMathOperator{\Var}{Var}
\DeclareMathOperator{\Li}{Li}

\hypersetup{
    colorlinks = true,
    linkcolor  = blue,
    citecolor  = blue,
    urlcolor   = blue,
}
\begin{document}

\title{From compact localized states to many-body scars in the random quantum comb}
 \author{Oliver Hart\,\href{https://orcid.org/0000-0002-5391-7483}{\includegraphics[width=6.5pt]{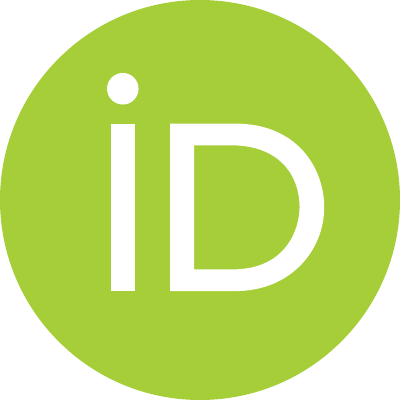}}}
 \affiliation{T.C.M.~Group, Cavendish Laboratory,  JJ~Thomson Avenue, Cambridge CB3 0HE, United Kingdom}
 \author{Giuseppe De Tomasi}
 \affiliation{T.C.M.~Group, Cavendish Laboratory,  JJ~Thomson Avenue, Cambridge CB3 0HE, United Kingdom}
 \author{Claudio Castelnovo}
 \affiliation{T.C.M.~Group, Cavendish Laboratory,  JJ~Thomson Avenue, Cambridge CB3 0HE, United Kingdom}

 \date{May 2020}

\begin{abstract}
    \setstretch{1.05}
    In this work we investigate the effects of configurational disorder on the eigenstates and dynamical properties of a tight-binding model on a quasi-one-dimensional comb lattice,
    consisting of a backbone decorated with linear offshoots of randomly distributed lengths.
    We show that all eigenstates are exponentially localized along the backbone of the comb.
    Moreover,
    we demonstrate the presence of an extensive number of compact localized states with precisely zero localization length. 
    We provide an analytical understanding of these states and show that they survive in the presence of density-density interactions along the backbone of the system where, for sufficiently low but finite particle densities, they form many-body scar states.
    Finally, we discuss the implications of these compact localized states on the dynamical properties of systems with configurational disorder, and the corresponding appearance of long-lived transient behaviour in the time evolution of
    physically relevant product states.
\end{abstract}
%
%

\maketitle
\section{Introduction}

The role of disorder in quantum systems with itinerant degrees of freedom has been the subject of much interest over several decades since the pioneering work of Anderson~\cite{Anderson1958} and collaborators~\cite{Abrahams1979}.
The field has recently received renewed interest with a specific focus on the interplay between disorder and interactions in many-body localization~\cite{Basko2006,Pal_2010,Oganesyan2007,Nandki_MBL_15,Vasseur2016}, particularly with reference to questions about ergodicity breaking in closed quantum systems and the eigenstate thermalisation hypothesis~\cite{Deutsch1991,Srednicki1994,Rigol2008,Rigol_review_2016}.

Disorder can take many forms, and in some contexts it can be due to the structure or configuration of the system rather than random potential energy or random interaction terms in the Hamiltonian.
For example, an imperfect (quasi-) one-dimensional (1D) system may have dendritic offshoots~\cite{Mendez2013},
as in 
biological systems~\cite{H_fling_2013}, percolation clusters~\cite{Harris1982}, and spin chains~\cite{Agrawal2019, White2020, Fulde95, Gong2010,Zaliznyak07,Chepiga19}.
Another example is that of quasi-particles in dimer, vertex, and ice models -- spin ice materials being a case in point~\cite{TomaselloPRL}; 
in these systems, spin correlations impose local constraints on the quasi-particles' dynamics that can result in quasi-1D structures with a backbone and a distribution of 
dangling ends or offshoots~\cite{SternSarang2019,HartVison2020}.

Here we take a closer look at this phenomenon by investigating a model system consisting of a 1D tight-binding chain with linear offshoots whose lengths are distributed randomly
(i.e., a random quantum comb model).
Classical variants of comb models have been introduced as a stepping stone to understanding diffusion in percolation clusters and other fractal systems~\cite{BOUCHAUD_90}; 
the offshoots lead to the trapping of particles, which inhibits their motion along the backbone, and can result in subdiffusive behaviour~\cite{White1984, Havlin1987, Pottier1994, Iomin2011, Villamaina2011, Lenzi2013, Agliari2014, Benichou2015, Yuste2016, Quintanilla2007}.

In the quantum model, we demonstrate that integrating out the offshoots of random lengths results in a localizing disorder for the quasi-particles moving along the backbone.
The configurational nature of the disorder produces resonances between the total energy of the particle and the energy levels of the offshoots.
These in turn give rise to an extensive number of states with vanishing localization length along the backbone, namely \emph{compact localized} (CL) states~\cite{Flach2014, Maimaiti2017_CLS}, see also Refs.~\cite{Danieli2020I,Danieli2020II,Danieli2020manybody,Kuno2020}.
We provide an analytical understanding of these states and a numerical study of their effect on the dynamics along the backbone.
Furthermore, we give conditions for their existence in comb-like structures with generic offshoots.
Importantly, these states appear also in specific translationally-invariant configurations of the comb structure, where they form flat bands.

We demonstrate that these CL states survive the addition of non-integrable interactions along the backbone.
They correspond to atypical, area law states in an otherwise thermal spectrum, i.e., quantum many-body scars~\cite{Kormos2017,Vafek2017Entanglement,Moudgalya_2018_scars1,Moudgalya_2018_scars,Bernien2017,Turner2018Nat,Turner2018,Shiraishi2017,Ho_2019_scars,Lin2019scars} that give rise to weak ergodicity breaking.
We argue that these scar states have a strong overlap with several physical states of interest in these systems, and are likely to give rise to long-lived transients in their dynamical properties before thermalization eventually sets in, as is expected asymptotically.

\begin{figure}[t]
    \includegraphics[width=0.7\linewidth]{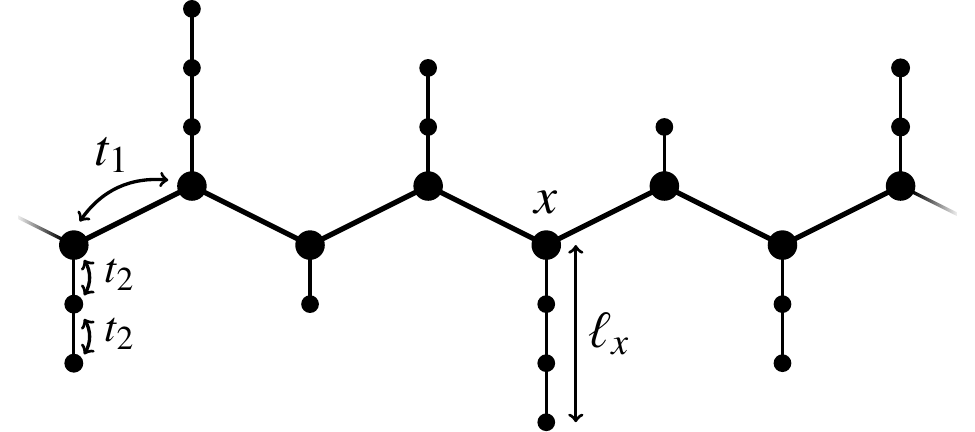}
    \caption{Diagram of a section of the random quantum comb system. The hopping parameter is $t_1$ in the backbone, and $t_2$ in the offshoots. The length $\ell_x$ of the linear offshoot at site $x$ on the backbone is drawn from the probability distribution $p(\ell)$.}
    \label{fig:comb}
\end{figure}

Our results were obtained by studying a model system to achieve an analytical understanding as well as allowing for numerical simulations of reasonably large systems and times.
However, the key properties that we uncovered ultimately hinge on a simple phenomenon: Integrating out the offshoots gives rise to an energy-dependent effective disorder that leads to resonances between the particle energy and the energy of the available states at each site.
We expect this behaviour to feature in a broader class of systems. Indeed, we find that the density of states in our model exhibits sharp features reminiscent of the ones observed in a recent study of quantum spin ice~\cite{SternSarang2019}, alluding to the possibility that such compact localized states may indeed play a role in the properties of that system.

The manuscript is structured as follows.
We first introduce the random quantum comb model in Sec.~\ref{sec:model}.
Then, in Sec.~\ref{sec:localization-properties}, we show that all eigenstates of the model are exponentially localized along the backbone, and look at the implications of this on the dynamics of the system. 
We discuss the compact localized states in Sec.~\ref{sec:compact-localized-states} and finally, in Sec.~\ref{sec:many-body-scars}, we add density-density interactions to the backbone of the comb and show that the aforementioned CL states form exact many-body scars.
We draw our conclusions and outlook in Sec.~\ref{sec:conclusions}.


\section{Model}
\label{sec:model}

We consider a single quantum particle hopping on a random,
comb-like structure (as shown in Fig.~\ref{fig:comb}) defined
by the Hamiltonian 
\begin{equation}
    \label{eq:H_comb}
    \hat{H}_0 = \hat{H}_{\text{B}} + \hat{H}_{\text{O}} + \hat{H}_{\text{B--O}}
    \, ,
\end{equation}
where
\begin{equation}
   \hat{H}_{\text{B}} = -t_1 \sum_{x=0}^{L-1} \left( \hat{c}_{x, 0}^\dagger \hat{c}_{x+1, 0}^{\phantom{\dagger}} + \text{H.c.} \right),
\end{equation}
describes hopping along the one-dimensional backbone and
\begin{equation}
   \hat{H}_{\text{O}} = - t_2 \sum_{x=0}^{L-1} \sum_{i_x = 1}^{\ell_x-1} \left( \hat{c}_{x, i_x}^\dagger \hat{c}_{x, i_x+1}^{\phantom{\dagger}} + \text{H.c.} \right),
\end{equation}
is the Hamiltonian on the offshoots, which take the form of one-dimensional chains of varying lengths.
The coupling between the two Hamiltonians is given by $\hat{H}_{\text{B--O}}=-t_2 \sum_x (\hat{c}_{x, 0}^\dagger \hat{c}_{x, 1}^{\phantom{\dagger}} + \text{H.c.} ) $.
The index $x$ labels the sites on the backbone, which satisfy periodic boundary conditions, and the indices $\{ i_x \}$ label the sites on the offshoots.
$L$ and $\ell_x$ are the lengths of the backbone and the offshoot on site $x$, respectively, in units of the lattice spacing.  
The randomness in the structure derives from the lengths of the offshoots, which
are drawn from a
probability distribution $p(\ell)$ with $\ell \in \mathbb{N}$.
The hopping amplitudes $t_1$ and $t_2$, corresponding to the backbone and the offshoots, respectively, are real and positive.
For $t_1=t_2$, Eq.~\eqref{eq:H_comb} represents the quantum version of the random comb model studied in Refs.~\cite{Pottier1994, Havlin1987, Yuste2016}.


\section{Localization properties}
\label{sec:localization-properties}

In this section, we prove analytically that all the eigenstates of $\hat{H}_0$ in Eq.~\eqref{eq:H_comb} are exponentially localized along the direction of the backbone by investigating the localization length $\xi_{\text{loc}}(E)$ at energy $E$ using transfer matrix techniques. We then study numerically the effects of these localized states on the dynamics of the system along the backbone.
%
%

\subsection{Transfer matrix results}
\label{sec:transfer-matrix}

\begin{figure}[t]
    \centering
    \includegraphics[width=1.\linewidth]{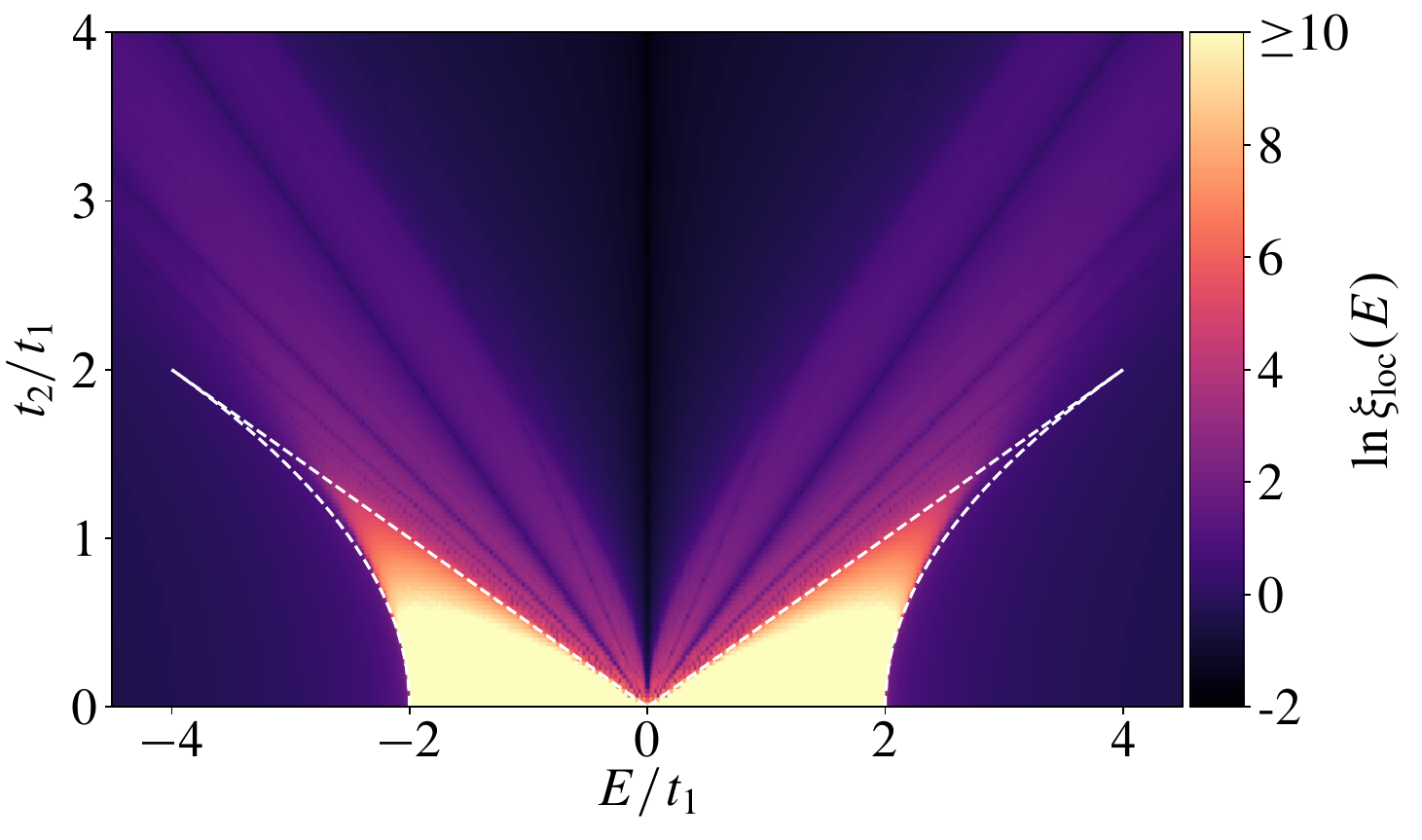}
    \caption{Localization length $\xi_\text{loc}$ as a function of energy $E$ and the ratio of the hopping parameters $t_2/t_1$ for a power-law probability distribution of lengths $p(\ell) \sim \ell^{-\gamma}$ with $\gamma = 2.5$, calculated using the transfer matrix technique, for systems of size $L \in [2\times10^5, 2\times 10^6]$ sites. The minima in $\xi_\text{loc}$ correspond to the discrete energy levels of the offshoots, $E \in \{E_n^{\text{cl}}\}$, which lead to a resonance in the magnitude of the effective disorder. The dashed region, $2t_2 < |E| < 2t_1 + \tfrac12 t_2^2/t_1$, contains no resonances.}
    \label{fig:loc-length}
\end{figure}

Denoting the projection of the wave function onto the sites with indices $(x, i_x)$ by $\psi_{x, i_x}$, the discrete form of the Schr\"{o}dinger equation according to the nearest-neighbour tight-binding Hamiltonian in Eq.~\eqref{eq:H_comb} is given by
\begin{equation}
    \label{eq:eq_backbone}
    -t_1 \psi_{x-1,0} - t_1 \psi_{x+1, 0} - t_2 \psi_{x, 1} = E \psi_{x, 0}
    \, ,
\end{equation}
for sites belonging to the backbone.
Further, for sites belonging to the offshoots, we can write down
\begin{subequations}
\begin{align}
    -t_2 \psi_{x, 0} -t_2 \psi_{x, 2} &= E \psi_{x, 1} \label{eq:Eq_offshoot_first} \\
    &\vdots \nonumber  \\
    -t_2 \psi_{x, \ell_x-1} &= E \psi_{x, \ell_x}
    \, ,
    \label{eq:Eq_offshoot}
\end{align}
\end{subequations}
where the length of the offshoot on site $x$ is $\ell_x$.
This set of equations, determining the wave function on the offshoots, can be solved recursively to give
\begin{equation}
    \label{eq:Eq_offshoot1}
    t_2\psi_{x, 1} = W_\epsilon(\ell_x) \psi_{x, 0}
    \, ,
\end{equation}
where we defined
\begin{equation}
    W_\epsilon(\ell) = -\frac{t_2}{\epsilon + \sqrt{\epsilon^2 - 1} \left[ 1 +  2\left( \left( \frac{\epsilon+\sqrt{ \epsilon^2 - 1 }}{ \epsilon-\sqrt{ \epsilon^2 - 1 } }\right)^{\ell}  -1\right)^{-1} \right] }
    \, ,
    \label{eqn:1d-offshoot-greens-fn}
\end{equation}
with $\epsilon \equiv E/2t_2$ the energy of the particle in units of half the bandwidth of the offshoots.
We further define $W_\epsilon(0) = 0$ for consistency of notation. Parametrising the energy
as $\epsilon = -\cos k$, we arrive at the more succinct expression
\begin{equation}
	W_\epsilon(\ell) = \frac{t_2 \sin(\ell k)}{\sin[(\ell + 1)k]}
	\, .
	\label{eqn:effective-disorder-simplified}
\end{equation}
For energies outside of the bandwidth of the offshoots, $|\epsilon| > 1$, the wave vector $k$ becomes
complex.

Equation~\eqref{eq:Eq_offshoot1} relates the wave function on the backbone $\psi_{x, 0}$ to the offshoot $\psi_{x, 1}$ connected to the same site.
Substituting Eq.~\eqref{eq:Eq_offshoot1} into the Schr\"{o}dinger equation on the backbone in Eq.~\eqref{eq:eq_backbone}, we obtain
\begin{equation}
    \label{eq:eq_effective}
    -t_1 \psi_{x-1,0} - t_1 \psi_{x+1, 0} = [E + W_\epsilon(\ell_x)] \psi_{x, 0}
    \, .
\end{equation}
Equation~\eqref{eq:eq_effective} describes an effective one-dimensional Anderson model with a random on-site potential energy term $\sum_x \mu_x \hat{c}^\dagger_{x,0} \hat{c}_{x,0}^{\phantom{\dagger}}$, with $\mu_x = - W_\epsilon(\ell_x)$, the magnitude of which depends explicitly on the energy $E$ of the particle.
Indeed, this process may be thought of as integrating out the offshoots to provide the sites on the backbone with a random self-energy.
The explicit energy dependence induced by the offshoots can result in curious dynamics along the backbone.
In the case of infinite offshoots attached to each site~\cite{Burioni2001}, the backbone dynamics can be described by a fractional time Schr\"{o}dinger equation~\cite{Iomin2011fractional,Iomin2019Markovian} (see also Appendix~\ref{sec:infinite-ofshoots}).

Crucially, the function $W_\epsilon(\ell)$ has $\ell$ simple poles.
As a result, the magnitude of the effective disorder diverges at these resonant energies, leading to a vanishing localization length as a function of energy, and the existence of CL states~\footnote{Note however that although in our model the two are in one-to-one correspondence, the existence of poles in the effective disorder does not generally imply the existence of compact localized states. See Appendix~\ref{sec:generic-offshoots} for further details.}.
Such states are also known to occur in percolation clusters in higher dimensions~\cite{Chayes1986}, and random fractal lattices~\cite{Kosior2017}.

In our case, the poles of $W_\epsilon(\ell)$ are at the energies of a chain of length $\ell$ with open boundary conditions, which has energy levels $E_n^\text{cl}(\ell) = -2t_2 \cos\left( \frac{n\pi}{\ell+1} \right)$, where $n = 1, \ldots, \ell$.
More generally, in the case of an arbitrary (non-interacting) offshoot with Hamiltonian $\hat{H}_x$ connected to the backbone at site $x$, the function $W_\epsilon$ is simply proportional to the diagonal element of the offshoot Green's function $\hat{G}_x(\epsilon) = (\hat{H}_x-\epsilon)^{-1}$ on the site connecting the offshoot to the backbone (see Appendix~\ref{sec:generic-offshoots} for further details).
The Green's function will then exhibit poles at the single-particle energies of that offshoot, i.e., the eigenvalues of the offshoot Hamiltonian $\hat{H}_x$.

\begin{figure}%
    \centering%
    \includegraphics[width=1.\linewidth]{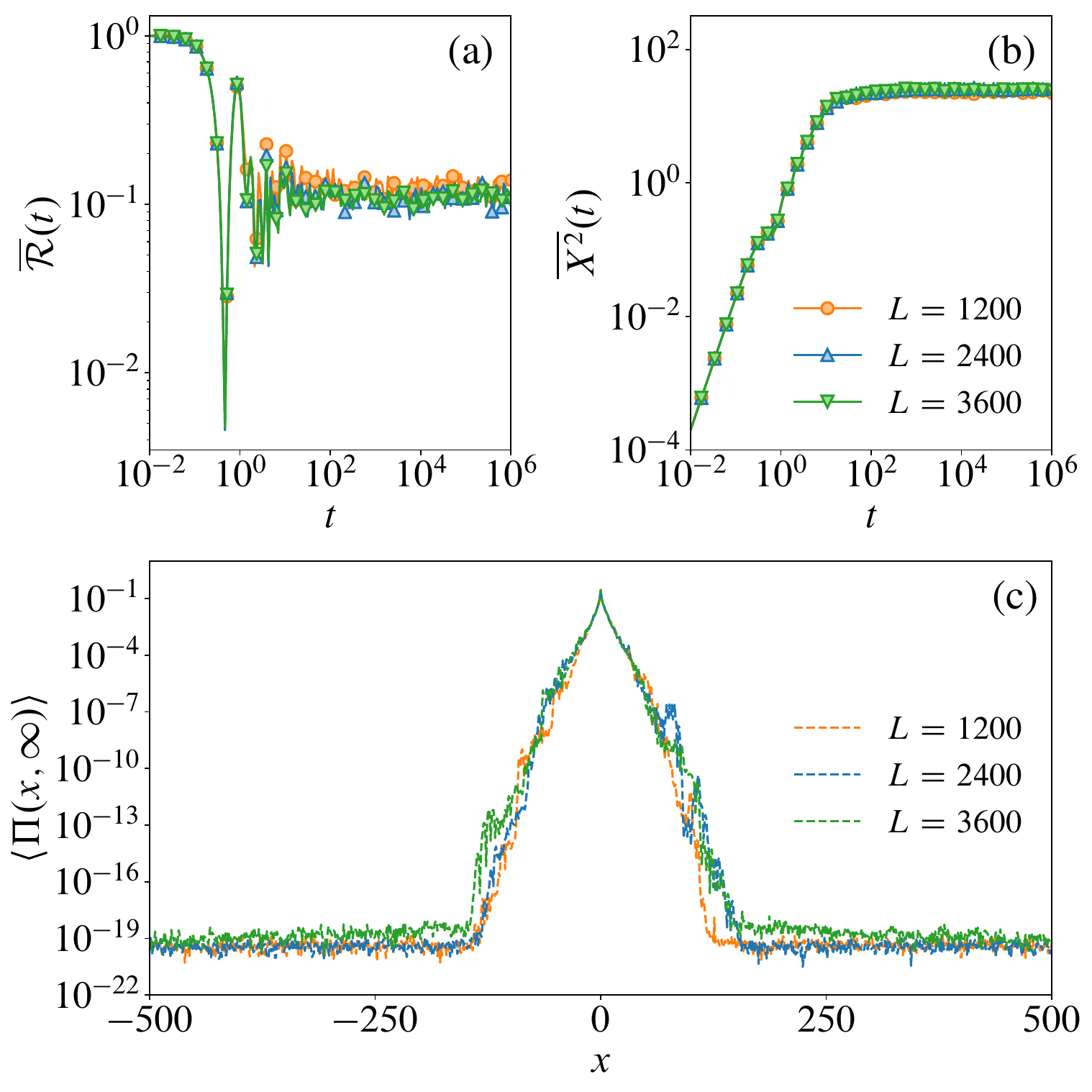}%
    \caption{Dynamics of the non-interacting system after starting with a wave packet initially localized on the site $x=0$ on the backbone. (a): Disorder-averaged return probability $\overline{\mathcal{R}}(t)$ for several system sizes $L\in \{1200, 2400, 3600\}$. (b): Disorder-averaged mean-square displacement $\overline{X^2}(t)$. (c): Long-time averaged probability distribution $\langle \Pi(x,\infty) \rangle$. In all panels, the probability distribution for the lengths of the offshoots is $p(\ell) \sim \ell^{-\gamma}$ with $\gamma =2.5$, and $t_2/t_1=\phi$ with $\phi = (1+\sqrt{5})$. }%
    \label{fig:Comb_loc_dynamic}%
\end{figure}%

For $E \ne \{E_n^{\text{cl}}\}$, i.e., irrational wave vectors $k/\pi \in \mathbb{R} \setminus \mathbb{Q}$ in Eq.~\eqref{eqn:effective-disorder-simplified}, we introduce the $2\times 2$ transfer matrices
\begin{equation}
    T_x(E)
    =
    \begin{pmatrix}
        -[E + W_\epsilon(\ell_x)]/t_1 & -1 \\
        1 & 0
    \end{pmatrix}
    \, ,
\end{equation}
which allow the Schr\"{o}dinger equation on the backbone, Eq.~\eqref{eq:eq_effective}, to be rewritten as
\begin{equation}
    \label{eq:Product_matrix}
    \begin{pmatrix}
        \psi_{x+1,0} \\
        \psi_{x,0}
    \end{pmatrix}
    =
    T_x(E) T_{x-1}(E) \cdots T_1(E) 
    \begin{pmatrix}
        \psi_{1,0} \\
        \psi_{0,0}
    \end{pmatrix}
    \, .
\end{equation}
The localization length at energy $E$ is then given by $\xi_\text{loc}(E) = \lambda^{-1}(E)$, where
\begin{equation}
    \lambda(E) = \lim_{L \to \infty}\frac{1}{L} \overline{ \ln || \tilde{T}_L (E)||}
    \, ,
\end{equation}
is the largest Lyapunov exponent of the product of transfer matrices $\tilde{T}_L(E) = \prod_{x=1}^L T_x(E)$ along the chain.
The overline indicates an average over the distribution of lengths of the offshoots.
Since $\det T_x = 1$, the product of transfer matrices $\tilde{T}_L$ also has unit determinant and hence its eigenvalues are reciprocals of one another.
For a nontrivial~\footnote{That is, excluding patterns of the offshoots that are translationally-invariant along the backbone direction.} probability distribution of the offshoots, $p(\ell)$, it is possible to use Furstenberg's theorem~\cite{Furstenberg1963} to show that the Lyapunov exponent is strictly positive and thus the localization length is finite $\xi_{\text{loc}}(E) < \infty$.

Figure~\ref{fig:loc-length} shows a colour plot of the localization length $\xi_{\text{loc}}$ as a function of the particle energy $E$ and the ratio $t_2/t_1$, where $\xi_{\text{loc}}$ is computed numerically using standard transfer matrix techniques~\cite{Kramer1993,crisanti2012products}. We checked numerically that the behaviour of $\xi_{\text{loc}}(E)$ changes only quantitatively, but not qualitatively, if we use a different probability distribution $p(\ell)$.
As already discussed, for energies satisfying the resonance condition $E \in \{E_n^{\text{cl}}\}$ the localization length drops to zero, as one can see from Fig.~\ref{fig:loc-length}.
This results in an intricate, fractal-like energy-dependence of the localization length for $|E| < 2 t_2$, vanishing for all rational wave vectors $k / \pi \in \mathbb{Q}$.
It is interesting to note that $\xi_{\text{loc}}(E)$ is largest at the edges of the energy spectrum, while in a standard Anderson model the localization length is instead largest for energies belonging to the middle of the spectrum~\cite{Kramer1993}.
This behaviour may be understood by noting that the effective disordered potential $W_\epsilon(\ell_x)$ is smallest at the edges of the spectrum, implying a larger localization length.
For example, in the simplest case of a binomial distribution of lengths, $p(\ell) = \frac{1}{2} [ \delta_{\ell,0} + \delta_{\ell,1} ]$, the fluctuations of the potential $W_\epsilon(\ell)$ are $ \Var[W_\epsilon(\ell)] = t_2^4/(2E^4)$, i.e., a decreasing function of $|E|$.
For $t_2/t_1 < 2$, there exist states (within the white dashed region in Fig.~\ref{fig:loc-length}) that are not in the vicinity of any of the offshoot eigenvalues $\{ E_n^{\text{cl}} \}$ (see Appendix~\ref{sec:bandwidth}).
Further, the offshoot Green's function decays with energy outside of the band $-2t_2 < E < 2t_2$, leading to a suppression of the disorder: $W_\epsilon(\ell) \simeq - t_2 e^{-k} (1 - 2e^{-2\ell k})$, where $\epsilon = \cosh k$. The localization length therefore becomes exponentially large in $k$ within this region.
%
%

\subsection{Dynamical properties}

We focus on the dynamics of the probability density $\Pi(x, t)$ starting from a wave packet localized on a single site of the chain, $|\psi(0)\rangle = \hat{c}_{0,0}^\dagger |0\rangle$.
In particular, we study the probability density marginalised over the offshoot indices
\begin{equation}
    \Pi(x,t) = \sum_{i_x=0}^{l_x} |\langle x, i_x |\psi(t) \rangle |^2
    \, ,
\end{equation}
that is, the probability of finding the particle on any site of the comb with backbone index $x$.
To quantify the spread of $\Pi(x,t)$, we define its return probability by $\mathcal{R}(t) = \Pi(x=0, t)$ and its mean-square displacement $X^2(t) = \sum_x x^2 \Pi(x,t)$.

\begin{figure}[t]
    \centering
    \includegraphics[width=1.\linewidth]{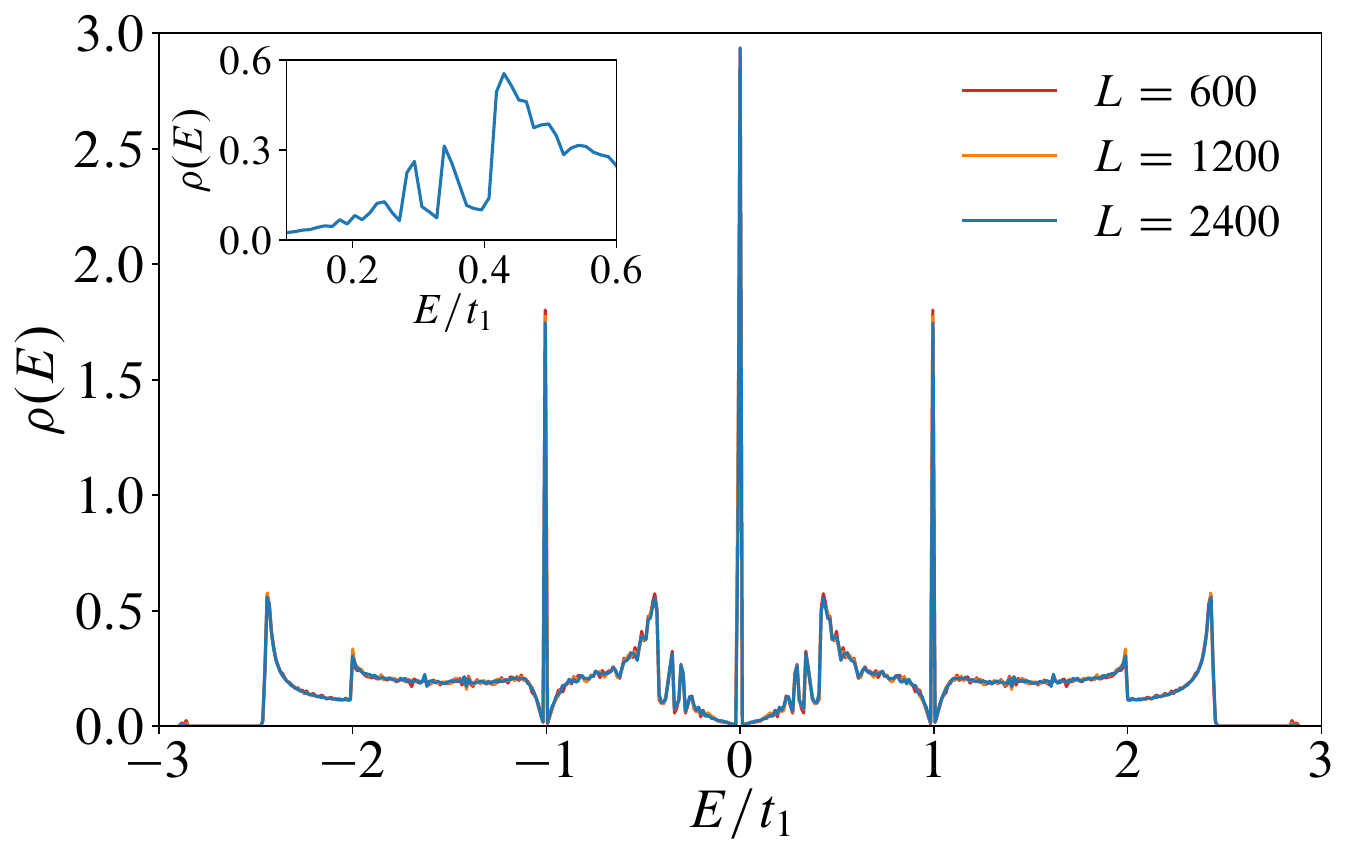}
    \caption{Configuration-averaged density of states $\rho(E)$ for several system sizes $L$ for the case in which the lengths of the offshoots are power law distributed $p(\ell)\sim \ell^{-\gamma}$ with $\gamma = 2.5$, and $t_2=t_1$. There exist non-analytic points at energies $E \in \{ E_n^{\text{cl}} \}$ that correspond to the CL states, a subset of which are shown in Eqs.~\eqref{eqn:121-zero-mode} and \eqref{eq:eigenstates_t2} that contribute towards the states at $E=0$ and $E=\pm t_2$, respectively. The inset shows $\rho(E)$ for $E/t_1\in [0,0.6]$ to highlight the presence of other non-analytic points.}
    \label{fig:density_states}
\end{figure}

As expected, and in agreement with the analysis of $\xi_{\text{loc}}(E)$, both the return probability $\overline{\mathcal{R}}(t)$ and the mean-square displacement $\overline{ X^2 } (t)$ exhibit behaviour typical of a localized system.
Figures~\ref{fig:Comb_loc_dynamic}(a) and \ref{fig:Comb_loc_dynamic}(b) show $\overline{\mathcal{R}}(t)$ and $\overline{ X^2 } (t)$ for several system sizes (i.e., lengths of the backbone $L$), where the dynamics has been computed using exact diagonalization.
The lengths of the offshoot are power-law distributed according to $p(\ell) \propto 1/\ell^{2.5}$~\footnote{In general, we define the power law distribution $p(\ell)$ with the inclusion of a cut-off equal to the length of the backbone, $L$. Thus,  $\ell \in \{1,...,L\}$ and $p(\ell) = \ell^{-\gamma}/\mathcal{N}$ with $\mathcal{N} = \sum_{k=1}^L k^{-\gamma}$.}, and $t_2/t_1=\phi$ with $\phi= 1+\sqrt{5}$. 
Both quantities, after some initial transient dynamics, saturate to an $L$-independent value, implying that the particle cannot propagate through the system beyond the maximal localization length.
Finally, Fig.~\ref{fig:Comb_loc_dynamic}~(c) shows the density profile averaged over both disorder realisations and time: $\langle \Pi(x, \infty) \rangle = \lim_{T\rightarrow \infty} \frac{1}{T} \int_0^T \overline{\Pi (x,t)} dt$, which relaxes to a stationary, exponentially decaying function $\langle \Pi(x, \infty) \rangle \sim e^{-|x|/\xi}$.

Interestingly, for $t_1=t_2$, where the maximum localization length becomes large, $\xi_\text{loc} \gg 1$ (see Appendix~\ref{sec:off-resonance-states}), we found a transient dynamics that is consistent with an algebraic propagation $\overline{X^2}(t) \sim t^\alpha$, where $\alpha \approx 2 - 1/\gamma$, with $\gamma > 2$ the decay rate of the power law probability distribution of the offshoots $p(\ell) \sim 1/\ell^\gamma$.
%
%


\section{Compact localized states}
\label{sec:compact-localized-states}

Having shown that all the eigenstates of $\hat{H}_0$ are exponentially localized along the backbone, we now turn our focus to the eigenstates of $\hat{H}_0$ with $\xi_{\text{loc}}(E)=0$, referred to previously as ``compact localized'' (CL) states.
As we already discussed, these states may be found at the zeros of the function $W^{-1}_\epsilon(\ell)$, which may be thought of as the inverse of the disordered onsite potential.

We are able to construct families of exact eigenstates of the full Hamiltonian, Eq.~\eqref{eq:H_comb}, for all values of $t_1/t_2$ at energies $E \in \{E_n^{\text{cl}}\}$
by considering symmetric clusters containing few sites.
For example,
\begin{equation}
    \raisebox{-0.3cm}{\includegraphics[width=1.15cm]{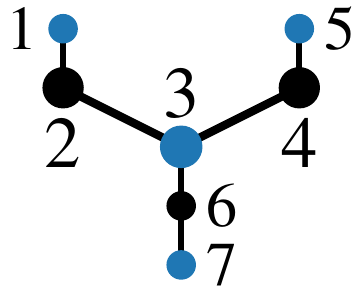}}
    \quad \text{has zero mode} \quad
    \left( \frac{t_1}{t_2}, 0, -1, 0, \frac{t_1}{t_2}, 0, 1 \right)^T
    \label{eqn:121-zero-mode}
\end{equation}
where the site labels correspond to their position in the state vector, and the blue sites correspond to the sites on which the wave function has a nonzero projection.
These states are in fact exact eigenstates
of the \emph{full} Hamiltonian in Eq.~\eqref{eq:H_comb} by virtue of having precisely zero projection onto the sites that
connect this cluster to the remainder of the lattice.
Further, placing any chain of even length on the intervening site will give rise to an eigenvector
of the form $(t_1/t_2, 0, -1, 0, t_1/t_2) \oplus (0, 1, 0, -1, \ldots)$ with zero energy.
\begin{figure}[t]
    \centering
    \includegraphics[width=0.75\linewidth]{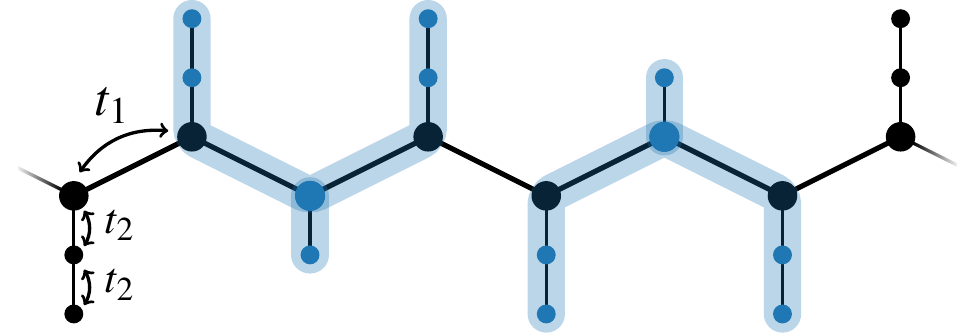}%
    \raisebox{-0.35cm}{\includegraphics[width=0.25\linewidth]{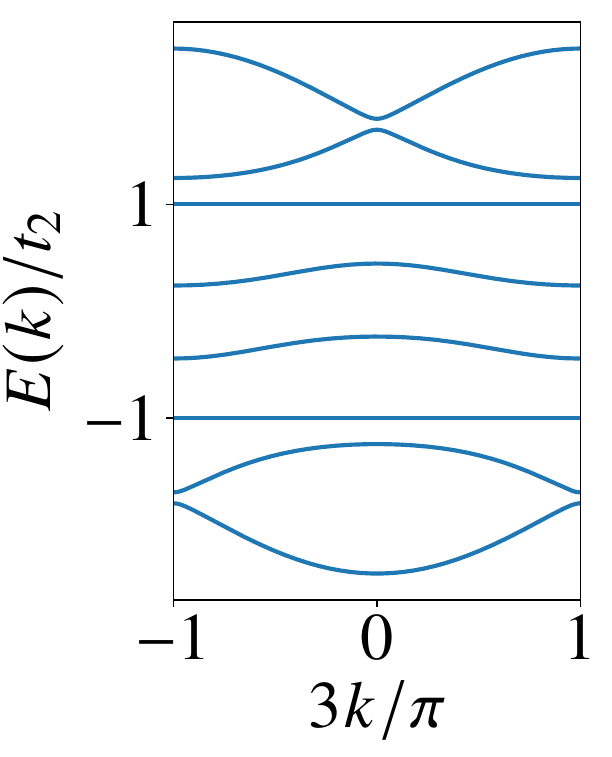}}%
    \caption{Translationally-invariant comb structure that hosts flat bands at $E = \pm t_2$ corresponding to the CL states discussed in Sec.~\ref{sec:compact-localized-states}.
    The support of the CL states is depicted by the blue sites.
    $t_1/t_2=1$ is chosen for the plot of the band structure, although the flat bands persist irrespective of the value of $t_1/t_2$.}
    \label{fig:simplified_system_212}
\end{figure}

Similarly, we are able to find clusters of sites that give rise to $\pm t_2$
eigenvalue pairs. For example,
\begin{align}
    \raisebox{-0.4cm}{\includegraphics[width=1.15cm]{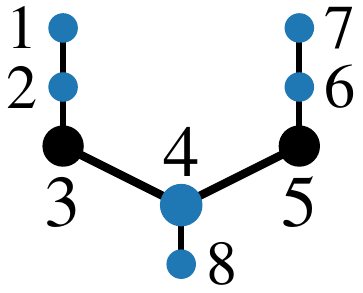}}
    \quad &\text{has two eigenstates with } E= \pm t_2 \nonumber \\
    \ket{\psi_\pm}=&\left(\frac{t_1}{t_2}, \mp \frac{t_1}{t_2}, 0, \pm 1, 0, \mp \frac{t_1}{t_2}, \frac{t_1}{t_2}, -1\right)^T \label{eq:eigenstates_t2}
    \, .
\end{align}
Once again, similar eigenstates with energy $\pm t_2$ may be written down for any intervening chain of length $\ell = 3m-2$ with $m \in \mathbb{N}$.

In the case of generic offshoots connected to the backbone,
the existence of such CL states is not guaranteed.
Indeed, using the above construction, if the two ends of the cluster at sites $x-1$ and $x+1$ have an identical offshoot with an eigenstate at energy $E_n^{\text{cl}}$,
then the intervening offshoot at site $x$ must satisfy $E_n^{\text{cl}} + W_{E_n^{\text{cl}}}(\ell_x) = 0$ for there to exist a CL state with energy $E_n^{\text{cl}}$ (i.e., an eigenstate satisfying $\psi_{x-1,0}=\psi_{x+1,0}=0$, which does not connect to the remainder of the chain).
This general condition, twinned with the exact expression for $W_\epsilon(\ell)$ in Eq.~\eqref{eqn:1d-offshoot-greens-fn}, can then be used to deduce the rule for the length of the intervening chain required to give rise to a CL state at energy $E_n^{\text{cl}}$:
\begin{equation}
    \ell_x = \frac{m(\ell_{x+1} +1)}{n} - 2
    \, ,
    \label{eqn:CL-length-requirement}
\end{equation}
with $m \subset \mathbb{N}$ such that $\ell_x \in \mathbb{N}$.
Such considerations also apply in higher dimensions (i.e., a clean $d$-dimensional hypercubic
tight-binding model augmented by linear offshoots), where analogous states can be constructed.
In this case, a compact localised state is hosted by an arrangement of offshoots satisfying Eq.~\eqref{eqn:CL-length-requirement}
where the central site is connected to an offshoot of length $\ell_x$, and all neighbouring sites are connected to offshoots of length
$\ell_{x+1}$.

It is important to note that these CL states are distributed with a finite density throughout the bandwidth of the offshoots $|E| < 2t_2$, and the corresponding eigenvalues, $E_n^{\text{cl}}(\ell) = -2t_2 \cos\left( \frac{n\pi}{\ell+1} \right)$, are extensively degenerate.
It is easy to estimate the average number of such states at $E \in \{E_n^{\text{cl}}\}$ by counting the expected number of occurrences of the above structures.
A consequence of these macroscopically degenerate energies is that the configuration-averaged density of states
$\rho(E) =  \overline{\sum_{\tilde{E}}\frac{\delta(E-\tilde{E})} {\text{dim}(\mathcal{H})}} $, where $\text{dim}(\mathcal{H})$ is the dimension of the Hilbert space,
will exhibit non-analytic points at $E \in \{E_n^{\text{cl}}\}$, as shown in Fig.~\ref{fig:density_states}.  
These CL states can also be found in special translationally invariant comb structures, where they form flat bands. Figure~\ref{fig:simplified_system_212} shows a comb lattice that hosts $2L/3$ such CL states at energies $E = \pm t_2$ [see Eq.~\eqref{eq:eigenstates_t2}].
%
%


\section{Many-body scars}
\label{sec:many-body-scars}

We now tackle the question of adding interactions to the quantum comb model in Eq.~\eqref{eq:H_comb}. We focus specifically on the fate of the CL states introduced in the previous section once density-density interactions are added between adjacent sites on the backbone.

Specifically, we consider the following deformation of the free Hamiltonian $\hat{H}_0$ defined in Eq.~\eqref{eq:H_comb}
\begin{equation}
    \hat{H} = \hat{H}_0 + V \hat{H}_\text{int}
    \, ,
    \label{eqn:H_int}
\end{equation}
where $\hat{H}_\text{int} = \sum_x \hat{n}_{x,0}  \hat{n}_{x+1,0}$, with ${\hat{n}}_{x,0} = {\hat{c}}^\dagger_{x,0} {\hat{c}}_{x,0}^{\phantom{\dagger}}$, corresponds to density-density interactions~\footnote{We use periodic boundary conditions.} on the backbone of magnitude $V$~\footnote{We note that the zero modes of the chain persist even when interactions are extended to include density-density interactions between adjacent sites on the offshoots.}.
It is easy to see that due to the spatial structure of the CL states, a Slater determinant of CL states belongs to the kernel of the interaction operator $\hat{H}_\text{int}$.
Specifically, let us define $\hat{\eta}^\dagger_{s}(E_n^{\text{cl}}) = \sum_x \sum_{i_x} \psi_{x,i_x}^{(s)}(E_{n}^{\text{cl}}) \hat{c}_{x,i_x}^\dagger$ as the creation operator for the single-particle CL state at energy $E_{n}^{\text{cl}}$, where the index $s$ labels its macroscopic degeneracy.
Now, non-interacting eigenstates of the form $\ket*{\psi_\text{cl}} = \prod_{n,s} \hat{\eta}_s^\dagger({E_n^{\text{cl}}}) |0\rangle$ remain eigenstates of $\hat{H}$ since they satisfy $\hat{H}_\text{int} \ket*{\psi_\text{cl}} = 0$.
One can only construct such states as long as the total number of particles does not exceed the total number of CL states.
Importantly, these eigenstates are highly non-thermal and violate the eigenstate thermalization hypothesis~\cite{Rigol_review_2016}, since they are exact eigenstates of a quadratic Hamiltonian.
Moreover, as a result of their strictly localized nature, they satisfy exact area law scaling of the entanglement entropy.

\begin{figure}[t]
    \centering
    \includegraphics[width=1.\linewidth]{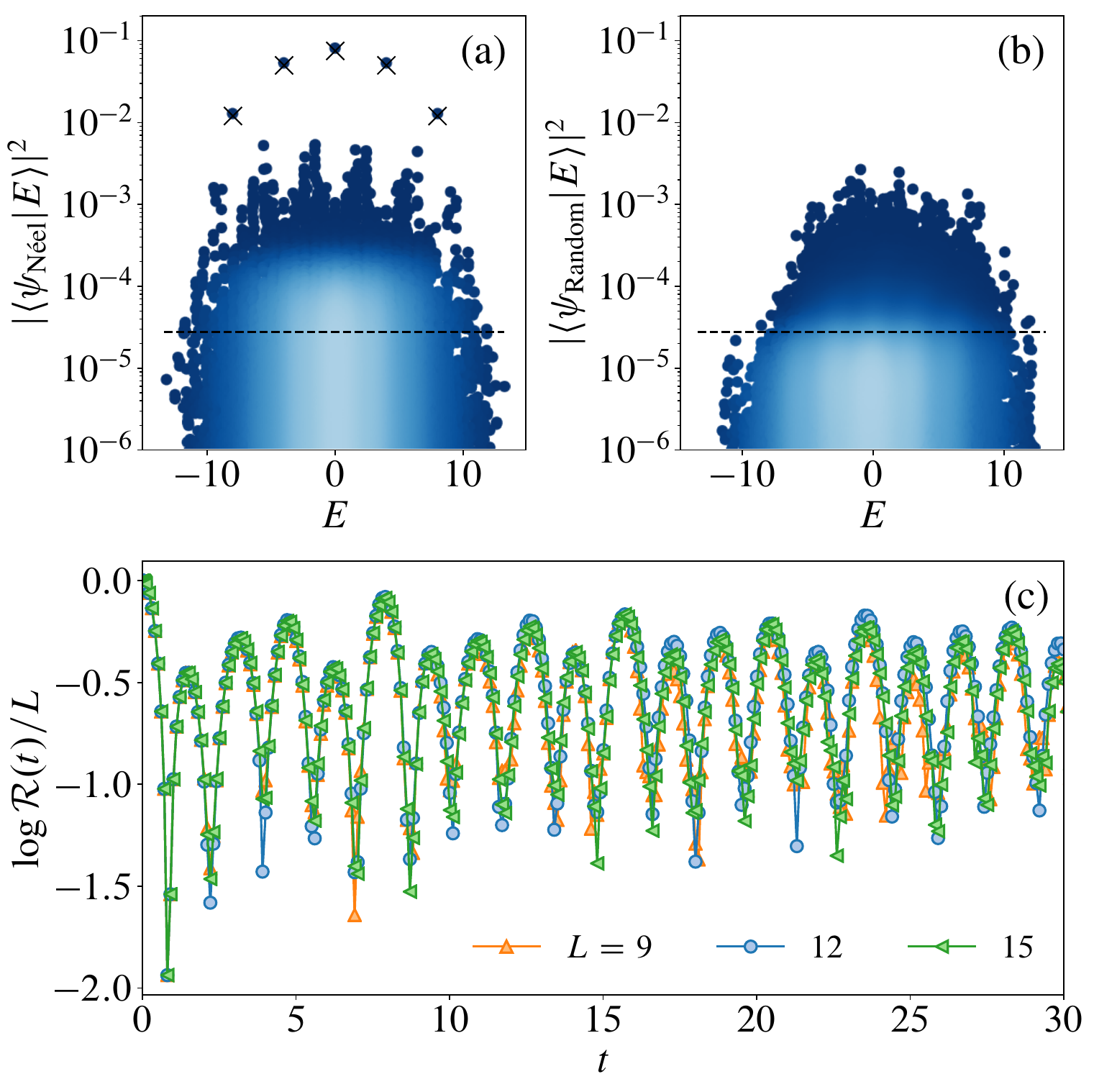}%
    \caption{(a): Projection of the N\'{e}el product state $|\psi_{\text{N\'{e}el}} \rangle =\prod_{x=1}^{L/3} \hat{c}^\dagger_{{3x-2},0}|0\rangle$ onto energy $E$ of the interacting Hamiltonian $\hat{H}$ in Eq.~\eqref{eqn:H_int} on the comb lattice in Fig.~\ref{fig:simplified_system_212} with $L=12$. The scar states can be seen at energies $E=0, \pm 2t_2, \pm 4t_2$ (indicated by the crosses)~\cite{overlap_footnote} (b): Projection of a random product state $\ket{\psi_{\text{Random}}}$ in the computational basis $\prod_{x, i_x} \hat{c}^\dagger_{x,i_x}|0\rangle$ onto energy $E$ of $\hat{H}$. 
    In both panels, the horizontal dashed lines represent the case of a state that is spread homogeneously over the Hilbert space $\sim \binom{8L/3}{L/3}^{-1}$ (fully ergodic), and the heatmap corresponds to the local density of points.
    (c): The return probability $\mathcal{R}(t) = |\langle \psi_{\text{N\'{e}el}} | e^{-i \hat{H}t} | \psi_{\text{N\'{e}el}} \rangle |^2$ for three system sizes $L = \{9, 12,15\}$ and interaction strength $V=1$. In all panels the number of particles is $N=L/3$ and $t_2/t_1 = 2$.}
    \label{fig:many_body}
\end{figure}

It is important to point out that the existence of these states does not depend on the value of the interaction strength $V$ or the hopping parameters $t_1$ and $t_2$.
Thus, in general, the integrability of the model is broken (see Appendix~\ref{sec:level-stats}), and hence these CL states are surrounded by thermal (volume law entangled) eigenstates.
These special states therefore constitute an example of exact many-body scar states~\cite{Kormos2017,Vafek2017Entanglement,Moudgalya_2018_scars1,Moudgalya_2018_scars,Bernien2017,Turner2018Nat,Turner2018,Shiraishi2017,Ho_2019_scars,Lin2019scars}.
Indeed, their construction is reminiscent of Refs.~\cite{Shiraishi2017,Choi2019,Ok2019}, for example.

As we already discussed for the non-interacting problem, these special states, located at some of the non-interacting energies, appear with probability one in random comb structures, as well as in special translationally invariant models.
For example, in the comb structure in Fig.~\ref{fig:simplified_system_212} with $N=L/3$ particles, we will have
$\binom{\frac{2L}{3}}{\frac{L}{3}}\sim 2^{2L/3}/\sqrt{\pi L/3}$
many-body scar states, which appear at energies $\pm n t_2$, with
$n\in \mathbb{Z}$.
Since these many-body scars are Slater determinants of CL states, it is easy to find physical (product) states in the computational basis, $\prod_{x, i_x} \hat{c}^\dagger_{x,i_x}|0\rangle$, that have a large overlap with them.
As a result, the dynamics starting from these special initial conditions will be strongly dominated by the existence of the many-body scars. Importantly, such initial conditions are relevant both theoretically as well as experimentally, e.g., in cold atom setups.

For concreteness, we focus on $\hat{H}$ defined on the translationally invariant comb structure shown in Fig.~\ref{fig:simplified_system_212}.
In this case, the charge density wave state $|\psi_{\text{N\'{e}el}}\rangle = \prod_{x=1}^{L/3} \hat{c}^\dagger_{{3x-2},0}|0\rangle$ maximises the overlap with the sub-Hilbert space spanned by the scar states.
Figure~\ref{fig:many_body}~(a) shows the overlap $|\langle\psi_{\text{N\'{e}el}}| E\rangle |^2$, between $|\psi_{\text{N\'{e}el}} \rangle$ 
and energy-resolved eigenstates of $\hat{H}$ for $L=12$, $N=L/3$, $t_2/t_1=2$ and $V=1$.
This shows that $|\psi_{\text{N\'{e}el}}\rangle$ is predominantly supported by the scar states, located at energies $E=0$, $\pm 2t_2$ and $\pm 4 t_2$.
It is important to note that $|\psi_{\text{N\'{e}el}}\rangle$ belongs to the middle of the spectrum of $\hat{H}$.
Conversely, starting from a random product state, i.e., infinite-temperature in the computational basis, the overlap $|\langle\psi_{\text{Random}}| E\rangle |^2$ is homogeneously spread over the entire energy spectrum of $\hat{H}$, as shown in Fig.~\ref{fig:many_body}~(b).

The presence of the scar states can be probed dynamically by studying the return probability $\mathcal{R}(t) = |\langle \psi_\text{N\'{e}el} | e^{-i \hat{H} t}| \psi_{\text{N\'{e}el}} \rangle |^2$.
We compute the time evolution of $\mathcal{R}(t)$ using Chebyshev polynomial techniques~\cite{Weiss06}, which is shown in Fig.~\ref{fig:many_body}~(c).
The return probability $\mathcal{R}(t)$ exhibits coherent oscillations, implying that the time-evolved state of the system is confined within the subspace spanned by the scar states, whose energies are commensurate.
Although the number of scar states is exponential in $L$, they occupy an exponentially small fraction of the total number of states in the thermodynamic limit.
The contribution to the time evolution of $\mathcal{R}(t)$ from the scar subspace may be calculated exactly:
\begin{equation}
    \frac{\log \mathcal{R}(t)}{L} \simeq \frac{1}{3} \log \cos^2\left( \frac{t}{\tau_2} \right) - \frac{2}{3} \log \left[ \frac{2 t_1^2 + t_2^2}{ t_2^2 } \right]
    \, ,
    \label{eqn:scar-return-prob}
\end{equation}
where $\tau_2 = t_2^{-1}$.
Since the CL states are exact eigenstates of the interacting Hamiltonian, the oscillations in~\eqref{eqn:scar-return-prob} will persist indefinitely.
%
%


\section{Conclusions}
\label{sec:conclusions}

In this work we studied tight-binding Hamiltonians on comb-like structures as a model system to investigate the effects of configurational disorder on localization and transport properties.
The model is composed of a one-dimensional backbone decorated with offshoots attached to each site of the backbone. These models are experimentally accessible in quantum synthetic platforms, such as cold atoms and other artificial systems.
Moreover, we argue that it may be relevant to the motion of quasi-particles in dimer and vertex models, quantum spin liquids, and fractonic systems, supported by recent results on quantum spin ice~\cite{TomaselloPRL, SternSarang2019}.

We focused primarily on the case in which the offshoots assume the form of one-dimensional chains whose lengths are randomly distributed.
Pictorially, this model represents a quasi-one-dimensional system in which a particle is able to escape from the main chain (backbone) due to the presence of the offshoots (see Fig.~\ref{fig:comb}).

We considered first the non-interacting limit of the model.
We showed analytically and numerically that all the eigenstates are 
exponentially localized along the direction of the backbone for any amount of disorder in the lengths of the offshoots.
Using transfer matrix techniques, we mapped the problem onto a one-dimensional Anderson model with an effective, energy-dependent, onsite disorder.
Analysing the behaviour of this effective onsite disorder, we identified special energies for which the onsite disorder diverges. As a result, at these energies, which coincide with the energy levels of the Hamiltonian of the offshoots, the eigenstates are compact localized (CL), characterised by a vanishing localization length. Moreover, the energy degeneracy of these states is extensive, leading to non-analytic points in the density of states.
These CL states are also present in translationally invariant comb structures, where they form flat bands.

Finally, we considered the interacting case, where the interactions act between adjacent sites on the backbone only. Analytically, we proved that Slater determinants of CL states are exact eigenstates of the interacting model, as long as the particle number is less than the number of such single-particle states. These states, which are highly non-thermal, have area law entanglement scaling while belonging to the bulk of the energy spectrum, providing an example of exactly solvable many-body scars.

Importantly, these states have a large overlap with experimentally relevant states. Therefore, we argue that for a range of physical initialisations of the system, the compact localized states alter dramatically the real-time evolution for experimentally accessible time scales.
As a result, we were able to provide a quench protocol to probe dynamically the existence of scar states in the system. Numerically, we tested it by investigating the dynamics of the interacting model in a translationally invariant comb structure, which hosts perfect quantum many-body scars.

Interesting directions for future work include investigating the possibility of a many-body localization~\cite{Nandki_MBL_15} transition in the interacting random comb model, or testing the robustness of the many-body scars uncovered in this work with respect to more generic types of interactions.

Further, comb-like structures can be used to study the dynamics of one-dimensional open quantum systems, where the offshoots play the role of a reservoir/sink for the particles.
This connection is related to the paradigm of \textit{fractional quantum mechanics}~\cite{Laskin_2000,Laskin_book_2018}.
For example, the quantum dynamics on the backbone of the non-interacting tight-binding model with infinite offshoots is described by a fractional time Schr\"{o}dinger equation that exhibits non-unitary time evolution~\cite{Iomin2011fractional,Iomin2019Markovian}.
Whether the fractional time Schr\"{o}dinger equation could be used to investigate interacting, open many-body systems has not been studied extensively so far and it
is an interesting question that deserves further attention in the future.

Whereas we considered here a model system to facilitate in-depth analytical and numerical understanding, the underlying mechanism at play is generic: Integrating out the offshoots produces an energy-dependent effective disorder that leads to resonances between the particle energy and the energy of the available states at each site. 
For example, the density of states in our model exhibits sharp features reminiscent of the ones observed in quantum spin ice~\cite{SternSarang2019}, suggesting that compact localized states may play a significant role in the dynamics of those systems and materials.


\begin{acknowledgements}
We are deeply indebted to Tom Gray for several insightful discussions throughout the course of this work from its inception.
We would also like to thank Pasquale Calabrese and Sarang Gopalakrishnan for useful discussions.
This work was supported in part by the Engineering and Physical Sciences Research Council (EPSRC) Grants No.~EP/P034616/1 and
No.~EP/M007065/1.
GDT acknowledges the hospitality of MPIPKS Dresden where part of the work was performed.
Some of the numerical simulations were performed using resources provided by the Cambridge Service for Data Driven Discovery (CSD3) operated by the University of Cambridge Research Computing Service (\href{www.csd3.cam.ac.uk}{www.csd3.cam.ac.uk}), provided by Dell EMC and Intel using Tier-2 funding from the Engineering and Physical Sciences Research Council (capital grant EP/P020259/1), and DiRAC funding from the Science and Technology Facilities Council (\href{www.dirac.ac.uk}{www.dirac.ac.uk}). 
\end{acknowledgements}


\appendix

\section{Generic offshoots}
\label{sec:generic-offshoots}

\begin{figure}
    \centering
    \includegraphics[width=0.3\linewidth]{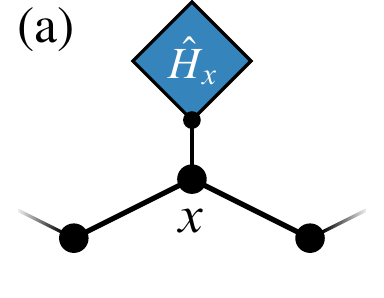}
    \includegraphics[width=0.5\linewidth]{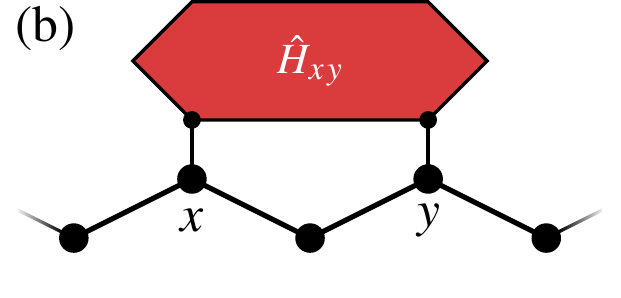}
    \caption{Depiction of generic offshoots emanating from the backbone. (a): An offshoot that may be integrated out to give a diagonal contribution to the disordered potential. (b): An offshoot that connects sites $x$ and $y$ on the backbone, leading to disorder in both the potential on sites $x$ and $y$, and a hopping term between the two sites.}
    \label{fig:generic-offshoots}
\end{figure}

In this appendix, we consider the case of generic offshoots emanating from---and possibly connecting---various sites on the backbone. Let us begin by considering an offshoot which, when isolated from the backbone, is described by a generic Hamiltonian $\hat{H}_x$, quadratic in fermionic operators.
If the offshoot is then connected to the backbone via a nearest-neighbour hopping term with amplitude $-t_2$, the relationship between the projection of the wave function onto $\ket{x, 0}$ and $\ket{x, 1}$ is given by
\begin{equation}
    \psi_{x, 1}(\omega) = t_2 \mel*{x, 1}{\hat{G}_x(\omega)}{x, 1} \psi_{x, 0}(\omega)
    \, ,
    \label{eqn:diagonal-greens-function}
\end{equation}
where we have defined the Green's function of the offshoot $\hat{G}_x(\omega) = (\hat{H}_x - \omega)^{-1}$.
Hence, the value of the effective disordered potential is proportional to the diagonal element of the Green's function on the site that connects the offshoot to the backbone.

If instead the offshoots connect to the backbone in multiple locations, then the offshoots not only induce effective on-site disorder, but also induce an effective disordered hopping term between the sites that are connected by the offshoot. For the case shown in Fig.~\ref{fig:generic-offshoots}~(b), where there exists an offshoot with (isolated) Hamiltonian $\hat{H}_{xy}$ that connects to the backbone at sites $x$ and $y$, then
\begin{equation}
    \begin{pmatrix}
        \psi_{x, 1} \\
        \psi_{y, 1}
    \end{pmatrix}
    = t_2
    \begin{pmatrix}
        G_{xx}(\omega) & G_{xy}(\omega) \\
        G_{yx}(\omega) & G_{yy}(\omega)
    \end{pmatrix}
    \begin{pmatrix}
        \psi_{x, 0} \\
        \psi_{y, 0}
    \end{pmatrix}
    \, ,
\end{equation}
where $G_{\alpha\beta}(\omega)\equiv\mel*{\alpha, 1}{\hat{G}_{xy}}{\beta, 1}$. The generalisation to offshoots that connect more than two sites on the backbone is simple: The matrix that connects the projection of the wave function onto the backbone and the connected offshoot sites is given by the appropriate sub-matrix of the offshoot Green's function.
%
%

\subsection{Compact localized states}

If there exist two identical offshoots with Hamiltonian $\hat{H}_{x \pm 1} = \hat{H}^{(1)}$ on sites $x \pm 1$, and an offshoot on the central site with Hamiltonian $\hat{H}_x = \hat{H}^{(0)}$, we look for the conditions imposed on the Hamiltonians $\hat{H}^{(i)}$ for there to exist a compact localized (CL) state on this cluster.
Again $\hat{H}^{(i)}$ correspond to generic noninteracting Hamiltonians, quadratic in fermionic operators.
The CL state (if it exists) corresponds to an eigenstate of the Hamiltonian $\hat{H}^{(1)}$ with energy $\omega_n$, which requires that $\psi_{x\pm 1, 0}=0$ (also required for the state to be an exact eigenstate of the full chain).
We then require that the remaining sites satisfy $\psi_{x+1,1}=\psi_{x-1,1}=-(t_1/t_2)\psi_{x,0}$.
The final requirement places restrictions on the Hamiltonian of the central offshoot. We find that $\omega_n \psi_{x, 0} = -t_2\psi_{x, 1}$ and, from Eq.~\eqref{eqn:diagonal-greens-function}, $\psi_{x, 1}=t_2 G^{(0)}(\omega_n)\psi_{x, 0}$, where the Green's function $G^{(0)}$ corresponds to the Hamiltonian $\hat{H}^{(0)}$. We therefore require the consistency condition
\begin{equation}
    t_2^2 G^{(0)}(\omega_n) + \omega_n \stackrel{!}{=} 0
    \, ,
    \label{eqn:compact-consistency-condition}
\end{equation}
to be satisfied in order for the CL state at $\omega_n$ to exist.
This equation determines which pairings of offshoots allow for the existence of CL states at the eigenvalues of $\hat{H}^{(1)}$.

For the case of linear offshoots, the appropriate diagonal element of the Green's function is given by Eq.~\eqref{eqn:1d-offshoot-greens-fn} in the main text. If the offshoots on sites $x\pm 1$ have length $\ell_{x+1}$, and the central offshoot has length $\ell_x$, then Eq.~\eqref{eqn:1d-offshoot-greens-fn} evaluates to
\begin{equation}
	\frac{W_{\epsilon_n}}{t_2} =  \frac{ \sin \ell_x k_n }{\sin [(\ell_x+1) k_n]  }
\end{equation}
for the energy $E_n = -2t_2 \cos k_n$.
Solving the consistency relation~\eqref{eqn:compact-consistency-condition} gives rise to the length constraint on the central offshoot stated in the main text: $\ell_x = (\ell_{x+1}+1)m/n - 2$, for $m \in \mathbb{N}$.

If the offshoots are instead given, for example, by Bethe branches with branching ratio $z-1$ (i.e., coordination number $z$), then the Green's function defined by Eq.~\eqref{eqn:1d-offshoot-greens-fn} is replaced by $G(E) \to G(E/\sqrt{z-1})/\sqrt{z-1}$, where $\ell$ is now interpreted as the depth of the tree (i.e., the maximum recursion depth).
If we repeat the above analysis to find the required depth of the interstitial offshoot, we find, in general, that the depth required to satisfy Eq.~\eqref{eqn:compact-consistency-condition} would be non-integer, $\ell_x \notin \mathbb{N}$.
The one exception is for zero modes, $E_n=0$, in which case the consistency condition can be trivially satisfied [$G(0)=0$] by having no offshoot on the intervening site.

These restrictions can be relaxed somewhat if we allow each site on the backbone to be connected to two (or more) offshoots.
For example, if two offshoots with identical Hamiltonians $\hat{H}_x$ are connected to the backbone at site $x$, then we can construct a CL state satisfying $\psi_{x,0}=0$ for each eigenstate of $\hat{H}_x$, which therefore does not connect to the rest of the backbone.
The CL state is formed by antisymmetrising an eigenstate $\ket{\varphi}$ of $\hat{H}_x$ so that the eigenstate of the full Hamiltonian is of the form $\ket{\varphi} \oplus 0 \oplus \ket{-\varphi}$.

\section{Off-resonance states}
\label{sec:off-resonance-states}

Here we provide an estimate of the scaling of the localization length in the region containing no resonances with the energy levels of the offshoots, i.e., the region surrounded by the dashed lines in Fig.~\ref{fig:loc-length}.
Outside of the energy band of the offshoots, the Green's function decays exponentially, leading to a suppressed disordered potential.
Defining $\epsilon = E/2t_2 = \cosh k$, the potential in this region equals 
\begin{equation}
    \frac{W_\epsilon(\ell)}{t_2} = \frac{1-e^{2\ell k}}{e^k e^{2\ell k} - e^{-k}}
    \, .
\end{equation}
If the energy lies outside of the bandwidth of the offshoots, $\epsilon > 1$, then we may approximate $W_\epsilon(\ell) \simeq - t_2 e^{-k}(1 - e^{-2\ell k})$. Within this approximation, we may then evaluate the variance of the disordered potential
\begin{align}
    \Var W_\epsilon &= e^{-k} \frac{t_2^4}{t_1^2} \left[ \frac{\Li_\gamma(e^{-4k})}{\zeta(\gamma)} - \left( \frac{\Li_\gamma (e^{-2k})}{\zeta(\gamma)} \right)^2 \right] \\
    &\simeq \frac{1}{(2\epsilon)^5} \frac{t_2^4}{t_1^2} \left[ \frac{1}{\zeta(\gamma)} - \frac{1}{\zeta^2(\gamma)} \right]
    \, ,
\end{align}
where $\Li_\gamma(x)$ is the Polylogarithm function, $\zeta(\gamma) = \Li_\gamma(1)$ is the zeta function, and $\gamma$ is the exponent in the power-law distribution of the offshoots lengths.
The approximate equality in the second line holds for $e^{-k} \ll 1$. This means that the system is ``most disordered'' for the power $\gamma \simeq 1.73$.

\section{Infinite offshoots}
\label{sec:infinite-ofshoots}

\subsection{Backbone dynamics}

\begin{figure}[t]
    \centering
    \includegraphics[width=\linewidth]{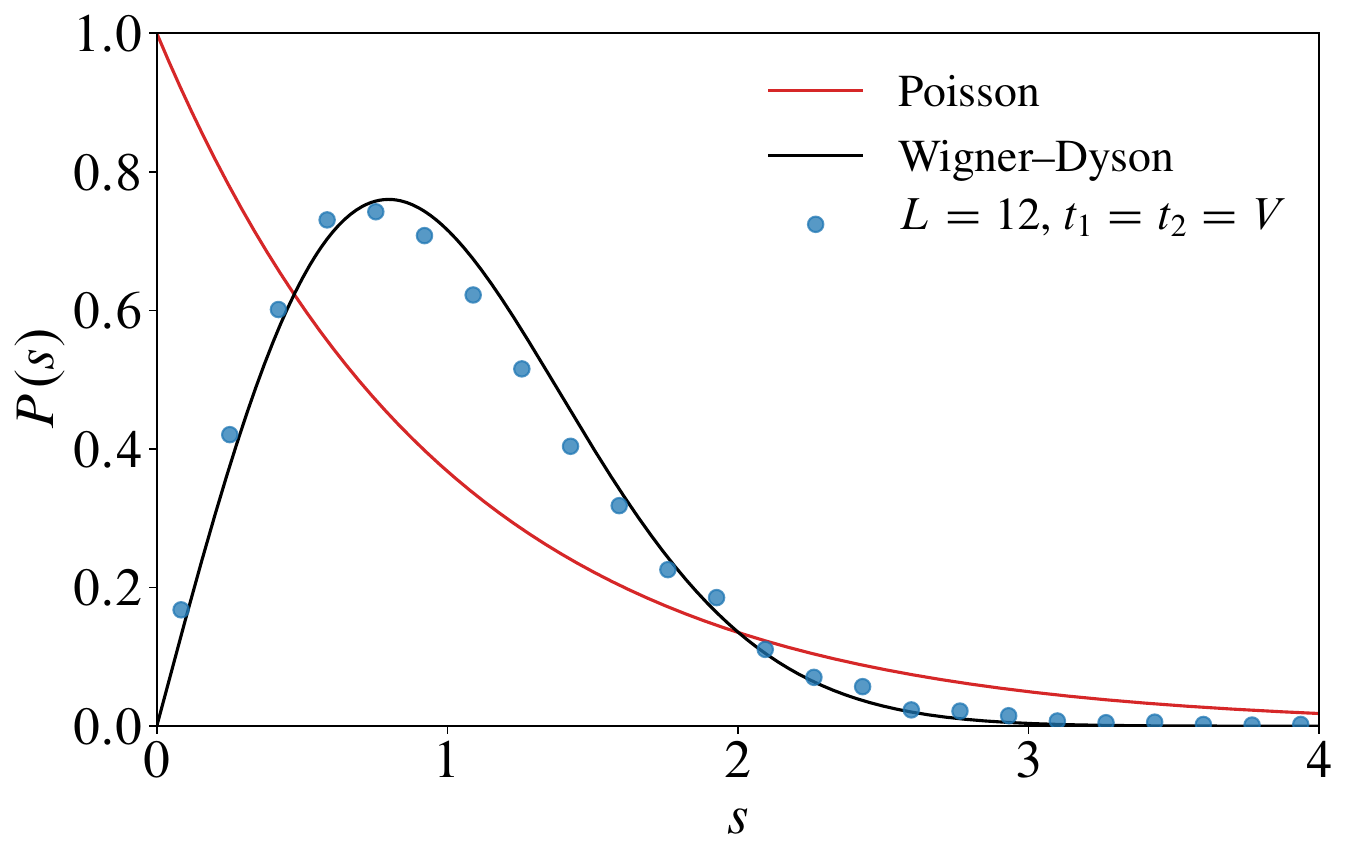}
    \caption{Level spacing distribution of the unfolded energy spectrum of the interacting Hamiltonian~\eqref{eqn:H_int} defined on the comb lattice shown in Fig.~\ref{fig:simplified_system_212} with $L=12$, and $N=L/3$ particles, having discarded $20\%$ of the energy levels at the edges of the spectrum. The distribution is averaged over momentum sectors, excluding $k=0, \pi$.}
    \label{fig:level-stats}
\end{figure}

In the case of infinite offshoots emanating from the backbone, the quantum comb still exhibits nontrivial dynamics when projected onto the backbone.
For convenience, let us impose periodic boundary conditions on the offshoots of length $N$ (including the backbone site).
Then, for a generic Hamiltonian $\hat{h}$ with matrix elements $h_{x, x'}$ along the backbone,
\begin{equation}
    \hat{H}_0 = \sum_{x, k_y} E(k_y) \hat{a}_{x,k_y}^\dagger \hat{a}_{x,k_y}^{\phantom{\dagger}} + \sum_{x,x'} h_{x,x'} \hat{c}_{x,0}^\dagger \hat{c}_{x',0}^{\phantom{\dagger}}
    \, ,
\end{equation}
where the dispersion for the offshoots is given by $E(k_y) = -2t_2 \cos k_y$, and $\hat{c}_{x,j}$ and $\hat{a}_{x,k_y}$ are related via Fourier transformation: $\hat{c}_{x, j} = N^{-1/2} \sum_{k_y} e^{i k_y j} \hat{a}_{x, k_y}$.
Let us denote the projection of the wave function onto the backbone on site $x$ by $\psi_{x,0}(t) = \bra{x,0}\ket{\psi}$, then the Schr\"{o}dinger equation $\hat{H}_0\ket{\psi} = i \partial_t \ket{\psi}$ can be written in momentum space as
\begin{equation}
    \tilde{\Psi}_{x, k_y}(s) = \frac{\frac{1}{\sqrt{N}}\sum_{x'} h_{x,x'}\Psi_{x', 0}(s) + i\tilde{\psi}_{x, k_y}(0)}{is + 2t_2 \cos k_y}
    \, ,
\end{equation}
where $\Psi_{x,j}(s) \equiv \mathcal{L}[\psi_{x,j}(t)]$ is the Laplace transform of $\psi_{x,j}(t)$, and $\tilde{\psi}_{x,k_y}$ denotes the (discrete) Fourier transform of $\psi_{x,j}$ over the direction of the offshoots.
Substituting this result back into the Schr\"{o}dinger equation and taking the sum over all momenta gives the dynamics of the wave function on the backbone:
\begin{equation}
    \frac{1}{N} \sum_{k_y}  \frac{i s [ \sum_{x'} h_{x,x'}\Psi_{x',0}(s) + i \psi_{x,0}(0) ]}{is - E(k_y)} = is \Psi_{x,0}(s)
    \, .
    \label{eqn:offshoots-integrated}
\end{equation}
This equation corresponds to nonunitary dynamics of the projection of the wave function onto the backbone, since probability density can be lost to the offshoots. Note that we have assumed that the particle begins on the backbone.

Intriguingly, in the continuum limit, where $E(k_y)=k_y^2/2m$, performing the integral over momentum and taking the inverse Laplace transform, one may write the result in terms of a \emph{fractional time} Schr\"{o}dinger equation of the form given in Ref.~\cite{Iomin2011fractional}, i.e., of the form $i^\alpha \partial_t^{\alpha} \psi = \mathcal{H}\psi$, where $\alpha = 1/2$.

\subsection{Bandwidth}
\label{sec:bandwidth}

In order to bound the bandwidth, we compute the spectrum of the translationally invariant model with infinite offshoots.
We will begin with offshoots of uniform length $N$, and take the $N \to \infty$ limit at the end of the calculation.
Taking the Fourier transform over the \emph{backbone} direction, we arrive at the Hamiltonian
\begin{equation}
    \hat{H}_0 = -2t_1\sum_{k_x} \cos k_x \hat{a}^\dagger_{k_x, 0} \hat{a}^{\phantom{\dagger}}_{k_x, 0} -t_2 \sum_{k_x, \langle i, j \rangle} \hat{a}^\dagger_{k_x, i} \hat{a}^{\phantom{\dagger}}_{k_x, j}
    \, .
\end{equation}
In particular, to find the bandwidth, we seek the extremal eigenvalues of $\hat{H}_0$ above.
Parametrising the energy as $E = -2t_2 \cosh \eta$, the quantisation condition for offshoots of length $N$ is found to be
\begin{equation}
    \frac{2t_1\cos k_x - t_2 e^{\eta}}{2t_1 \cos k_x - t_2 e^{-\eta}} = e^{-2N\eta}
    \, .
\end{equation}
In the thermodynamic limit $N \to \infty$, the solution of this equation for real $\eta$ is given by $2t_1 \cos k_x = t_2 e^\eta$.
This solution remains finite for $t_2 \to 0^+$, whilst the other $N-1$ eigenvalues vanish. Therefore, in the thermodynamic limit, we find the extremal eigenvalue for $t_2 < 2t_1$:
\begin{equation}
    |E| = 2t_1 + \frac{t_2^2}{2t_1}
    \, .
\end{equation}
This result defines the white dashed region in Fig.~\ref{fig:loc-length}.

\section{Level statistics}
\label{sec:level-stats}

To provide evidence that the system is non-integrable, we analyse the level statistics of the interacting Hamiltonian~\eqref{eqn:H_int}. In particular, we study the distribution of (unfolded) level spacings $P(s)$, where $s_n = E_{n+1} - E_n$~\cite{haake2001quantum}, and the level spacing statistics value  $\langle r_n \rangle = \langle \min(s_n, s_{n-1}) / \max(s_n, s_{n-1}) \rangle$~\cite{Oganesyan2007}.
It is crucial to resolve all symmetries of the system and calculate the statistics within each symmetry sector separately.
For the system shown in Fig.~\ref{fig:simplified_system_212}, the Hamiltonian possesses both translational invariance and inversion symmetry.
In Fig.~\ref{fig:level-stats}, we plot the level statistics for a system of size $L=12$, with $N=L/3$ particles averaged over momentum sectors (excluding $k=0,\pi$).
We find that the distribution $P(s)$ is in good agreement with the Gaussian orthogonal ensemble (GOE) from random matrix theory, as expected for non-integrable models~\cite{Rigol_review_2016}.
Further, the $r$-value is $\langle r \rangle = 0.534$, to be compared with the value of the GOE, $r_\text{GOE}=0.5359$, and of uncorrelated energy levels $r_\text{Poisson}=2\log{2}-1\approx0.3863$ for integrable models.


\bibliographystyle{aipnum4-1}
\bibliography{references}

\end{document}